\documentclass[runningheads,a4paper]{llncs}
\usepackage{llncsdoc}
\usepackage{breakurl}             
\usepackage{amssymb}
\usepackage{amsmath}
\usepackage{graphicx}
\usepackage{color}

\title{Distributed Synthesis of State-Dependent Switching Control}
\author{Adrien Le Co\"ent \inst{1}
\and
Laurent Fribourg \inst{2}
\and
Nicolas Markey \inst{2}
\and
Florian De Vuyst \inst{1}
\and
Ludovic Chamoin \inst{3}
}

\institute{CMLA, ENS Cachan, CNRS, Universit\'e Paris-Saclay \\
              61 av. du Pr\'esident Wilson, 94235 Cachan cedex \\
 	      \email{adrien.le-coent@ens-cachan.fr}
 	      \email{devuyst@cmla.ens-cachan.fr}
              \and
              LSV, ENS Cachan, CNRS, Universit\'e Paris-Saclay \\
              61 av. du Pr\'esident Wilson, 94235 Cachan cedex \\
               \email{fribourg,markey@lsv.ens-cachan.fr}
              \and
              LMT-Cachan, ENS Cachan, CNRS, Universit\'e Paris-Saclay \\
              61 av. du Pr\'esident Wilson, 94235 Cachan cedex \\
               \email{chamoin@lmt.ens-cachan.fr}
}

\titlerunning{Distributes synthesis of state-dependent switching control}
\authorrunning{A. Le Co\"ent, L. Fribourg, N. Markey, F. De Vuyst \& L. Chamoin}
 \begin{document}
\maketitle

\begin{abstract}
We present a correct-by-design method of state-dependent control synthesis 
for linear discrete-time switching systems. Given an objective region $R$ of the state space, 
the method builds a capture set $S$ and a control which steers 
any element of $S$ into $R$. The method works by iterated backward reachability 
from $R$. More precisely, $S$ is given as a parametric extension of $R$, and 
the maximum value of the parameter is solved by linear programming. 
The method can also be used to synthesize a stability control 
which maintains indefinitely within $R$
all the states starting at $R$.
We explain how the synthesis method can be performed in a distributed manner. 
The method has been implemented and successfully applied to the synthesis of 
a distributed control of a concrete floor heating system with $11$ rooms and 
$2^{11}=2048$ 
switching modes. 
\end{abstract}

\section{Introduction}
The importance of switched systems has grown up considerably these last years
in reason of their ease of implementation for controlling cyber-physical systems.
A switched system is a family of sub-systems,
each with its own dynamics characterized by a parameter mode $u$
whose values are in a finite set $U$
(see \cite{liberzon2012switching}). However, due to the composition
of many switched systems together, the global switched system
has a number of modes and dynamics which increases exponentially.
Take for example a heating system for a building of 11 rooms
(see \cite{larsen2015online}): each room $i$ has a  heater with 
2 mode values $\{${\em off,on}$\}$. This makes a combination
of $2^{11}=2048$ mode values. If we want to analyze the 
evolution of a trajectory on a horizon of $K$ units of discrete
time, we have to consider the dynamics corresponding to
$2^{11K}$ possible sequences of modes, which
is intractable even for small values of $K$. It is therefore essential
to design {\em compositional} methods in order to obtain control methods
of switched systems that give formal guarantees on the correct
behavior of the cyber physical systems.

In this paper, we give a symbolic compositional method which
allows to synthesize a control of linear discrete-time switched systems that is
guaranteed to satisfy  {\em attainability} and {\em stability}
properties.

The method starts from an objective region $R$ of the state space,
which is rectangular (i.e., is a product of closed intervals of reals).
It then generates in a backward manner, using linear programming techniques,
an increasing sequence of
nested rectangles $\{R^{(i)}\}_{i\geq 0}$ such that every trajectory
issued from $R^{(i)}$ is guaranteed to reach $R^{(i-1)}$ in a bounded
number of time units. Once $R^{(0)} = R$ is reached, 
the trajectory is also guaranteed to stay in $R$
indefinitely (stability).
The method relies on a simple operation of {\em tiling} of 
the rectangles $R^{(i)}$ in a finite number
of sub-rectangles (tiles), using a standard operation of {\em bisection}.
Although the method works in a backward fashion, it does not
require to inverse the linear dynamics of the system (via matrix inversion),
and does not compute {\em predecessors} of symbolic states (tiles),
but only {\em successors} using the forward dynamics.
This is useful in order to avoid numerical imprecisions,
especially when the dynamics are {\em contractive}, which happens
often in practical systems (see \cite{Mitchell07}).

Another contribution of this paper is a technique
of state {\em over-approximation} which allows 
a distributed control synthesis:
this over-approximation allows sub-system $1$ to infer a correct value for
its next local mode $u_1$ without knowing the exact value of the state
of sub-system $2$.
This distributed synthesis method is computationally efficient,
and works in presence of partial observability.
This is at the cost of the performance of the control
which usually makes the trajectories attain the objective
area in more steps than with a centralized approach.\\
\\ 
{\bf Related Work}\\
In symbolic analysis and control synthesis methods for hybrid systems, the method 
of backward reachability and the use of polyhedral symbolic states, as used here, 
is classical (see, e.g., \cite{asarin2000effective,gillula2011applications}). The use of tiling or partitioning 
the state-space using bisection is also classical (see, e.g.,
\cite{jaulinbook,girard2012low}).
The main original contribution of this paper is to give a simple technique 
of over-approximation, which allows one component to estimate 
the symbolic state of the other component, in presence of 
partial information. This is similar in spirit to an assume-guarantee 
reasoning where the controller synthesis for each sub-systems
assumes that some safety properties are are 
satisfied by the others \cite{alur1999reactive,meyer2015safety}.  
In contrast to \cite{fribourg2015games}, we do not 
need, for the mode selection of a sub-system, to explore
blindly all the possible mode choices made by the other sub-system.
This yields a drastic reduction of the complexity\footnote{This 
separability technique is made possible because the difference
equation 
$x_1(t+1)=f_1(x_1(t),x_2(t),u_1)$ (see Section \ref{ss:modes}) does not involve the
control mode $u_2$.}.
This approach allows us to treat a real case study 
which is intractable with a centralized approach.
This case study comes from \cite{larsen2015online}, and we use the same decomposition
of the system in two parts (rooms $1$-$5$ and rooms $6$-$11$).
In contrast to the work of \cite{larsen2015online} which uses an on-line and
heuristic approach with no formal guarantees, 
we use here an off-line formal method which
guarantees attainability and stability properties.\\
\\
%
%
%
%
{\bf Implementation}\\
The methods of control synthesis both in the centralized context
and in the distributed context have been implemented
in a prototype written in Octave \cite{octave}.
All the computation times
 given in the paper have been performed on a
2.80 GHz Intel Core i7-4810MQ CPU with 8 GB of memory.\\
\\
{\bf Plan}\\
The structure of this paper is as follows.
The class of systems considered and some preliminary definitions are given in 
Section \ref{sec:switch}. 
Our symbolic approach, which is based on the tiling of the state space
and backward reachability, is explained in Section
\ref{sec:tiling}.
In Section \ref{sec:one-step}, we present a centralized method to synthesize
a controller based on a ``generate-and-test'' tiling procedure.
A distributed approach is then given in Section \ref{sec:distr}
where we introduce a state over-approximation technique in order
to avoid the use of non-local information by
the subsystem controllers.
For both methods, we provide attainability and stability
guarantees on the controlled
trajectories  of the system. Finally, in Section 6, our distributed
approach is applied on a real-case study of temperature control in a building
with 11 rooms and $2^{11}=2048$ switching modes of control.

\section{State-dependent Switching Control}\label{sec:switch}
\subsection{Control modes}\label{ss:modes}
Consider the discrete-time system with {\em finite control}:
$$ x_1(t+1)=f_1(x_1(t),x_2(t),u_1)$$
$$x_2(t+1)=f_2(x_1(t),x_2(t),u_2)$$
where $x_1$ (resp. $x_2$) is the 1st-component (resp. 2nd-component)
of the state vector variable, which takes its values
in $\mathbb{R}^{n_1}$ (resp. $\mathbb{R}^{n_2}$), 
and $u_1$ (resp. $u_2$) is
the 1st (resp. 2nd) component of the control {\em mode} variable
which takes its values in the {\em finite} set $U_1$ (resp. $U_2$).
We will often use $x$ for $(x_1,x_2)$, $u$ for $(u_1,u_2)$,
and $n$ for $n_1+n_2$.
We will also abbreviate the set $U_1\times U_2$ as $U$.
Let $N$ be the cardinal of $U$, and $N_1$ (resp. $N_2$) the cardinal
of $U_1$ (resp. $U_2$). We have $N=N_1 \cdot N_2$.\\

More generally, we abbreviate the discrete-time system under
the form:
$$x(t+1)=f(x(t),u)$$
where $x$ is a vector state variable
which takes its values in 
$\mathbb{R}^n=\mathbb{R}^{n_1}\times \mathbb{R}^{n_2}$, 
$u$ is of the form $(u_1,u_2)$ where
$u_1$ takes its values in $U_1$ and $u_2$ in $U_2$.\\

In this context, we are interested by the following
{\em centralized} control synthesis problem: at  each discrete-time $t$, select the appropriate mode $u\in U$
in order to satisfy a given property. In this paper we focus on
{\em state-dependent} control, which means that, at each time $t$,
the selection of the value of $u$ is done by considering only the values of $x(t)$.

In the {\em distributed} context, the control synthesis problem consists
in selecting concurrently the value of $u_1$ in $U_1$ 
according to the value of $x_1(t)$ {\em only},
and the value of $u_2$ in $U_2$ according to the value of $x_2(t)$ {\em only}.\\

The properties that we consider are {\em attainability} properties:
given a set $S$ and a set $R$, we look for a control which
will steer any element of $S$ to $R$ in a bounded number of steps.
We will also consider {\em stability} properties, which means,
that once the state $x$ of the system is in $R$ at time $t$,
the control will maintain it in $R$ indefinitely at $t+1, t+2,\cdots$.
Actually, given a state set $R$, we will present a method which
does not start from a given set $S$, but {\em constructs} it,
together with a control which steers all the elements
of $S$ to $R$ within a bounded number of steps
($S$ can be seen as a ``capture set'' of $R$).\\

In this paper, we consider that $R$ and $S$ are ``rectangles''
of the state space.
More precisely, $R=R_1\times R_2$ is a rectangle of reals, i.e.,
$R$ is a product of $n$ closed intervals of reals,
and $R_1$ (resp. $R_2$) is a product of $n_1$ (resp. $n_2$) closed intervals
of reals.
Likewise, we assume that $S=S_1\times S_2$ is a rectangular
sub-area of the state space.


%
%
%
%
%


\begin{example}\label{ex:spec}
The centralized and distributed approaches will be illustrated by
the example of a two rooms apartment, heated by two heaters located 
in each room (adapted from (Girard)). In this example,
the objective is to control the temperature of the two rooms. There is heat exchange between the 
two rooms and with the environment. The {\em continuous}
dynamics of the system is given by the
equation:
 $$
 \dot{\left( \begin{matrix}T_1 \\ T_2 \end{matrix} \right)}=\left(\begin{matrix} - \alpha_{21} - \alpha_{e1}-\alpha_f{ u_1} & \alpha_{21} \\
 \alpha_{12} &-\alpha_{12}-\alpha_{e2} - \alpha_f u_2\end{matrix}
\right) 
\left( \begin{matrix}T_1 \\ T_2 \end{matrix} \right) + \left( \begin{matrix}
             \alpha_{e1} T_e + \alpha_f T_f { u_1} \\ \alpha_{e2}T_e + \alpha_fT_fu_2
            \end{matrix}\right).$$

Here $T_1$  and $T_2$ are the temperatures of the two rooms, 
and the state of the system corresponds to $T=(T_1,T_2)$.
The control mode variable $u_1$ (respectively $u_2$)
 can take the values $0$ or $1$ 
depending on whether the heater in room 1 (respectively room 2)
 is switched off or switched on (hence $U_1=U_2=\{0,1\}$).
Hence, here $n_1=n_2=1$, $N_1=N_2=2$ and $n=2,N=4$.

$T_e$ corresponds to the temperature of the 
 environment, and $T_f$ to the temperature of the heaters. 
 The values of the different parameters are the following: $\alpha_{12} = 5 \times 10^{-2}$,
 $\alpha_{21} = 5 \times 10^{-2}$, $\alpha_{e1} = 5 \times 10^{-3}$, $\alpha_{e2} = 5 \times 10^{-3}$,
 $\alpha_{f} = 8.3 \times 10^{-3}$, $T_e = 10$ and $T_f = 35$. 

We suppose that the heaters can be switched periodically
at sampling instants $\tau, 2\tau, \dots$ (here, $\tau=5s$).
By integration of the continuous dynamics between $t$ and $t+\tau$,
the system can be easily put under the 
desired {\em discrete-time} form: 

$T_1(t+1)=f_1(T_1(t),T_2(t),u_1)$

$T_2(t+1)=f_2(T_1(t),T_2(t),u_2)$, \\
where $f_1$ and $f_2$ are affine functions.

Given an objective rectangle for $T=(T_1,T_2)$ of the form
$R=\lbrack 18.5 , 22 \rbrack \times \lbrack 18.5 , 22 \rbrack$,
the control synthesis problem is to find a rectangular 
capture set $S$ 
as large as possible, from which one can
steer the state $T$ to $R$ (``attainability''), 
then maintain $T$ within $R$ for ever (``stability'').

\end{example}

\subsection{Control patterns}
It is often easier to design
a control of the system using several applications of $f$ in a row
rather than using just a single application of $f$ at each time.
We are thus led to the notion of ``macro-step'', and ``control pattern''.
A {\em (control) pattern $\pi=(\pi_1,\pi_2)$ of length $k$} 
is a sequence of modes defined recursively by:
\begin{enumerate}
\item $\pi$ is of the form $(u_1,u_2)\in U_1\times U_2$ if $k=1$,
\item $\pi$ is of the form $(u_1\cdot \pi'_1,u_2\cdot \pi'_2)$,
where $u_1$ (resp. $u_2$) is in $U_1$ (resp. $U_2$),
and $(\pi'_1,\pi'_2)$ is a (control) pattern of length $k-1$ if $k\geq 2$.
\end{enumerate}

The set of patterns of length $k$ is denoted by $\Pi^k$
(for length $k=1$, $\Pi^1=U$). Likewise, for $k\geq 1$,
we denote by $\Pi_1^k$ (resp. $\Pi_2^k$)
the set of sequences of $k$ elements of $U_1$ (resp. $U_2$).\\

For a system defined by $x(t+1)=f(x(t),(u_1,u_2))$
and a pattern $\pi=(\pi_1,\pi_2)$ of length $k$,
one can define recursively $x(t+k)=f(x(t),(\pi_1,\pi_2))$
with $(\pi_1,\pi_2)\in \Pi^k$,
by:
\begin{enumerate}
\item $f(x(t),(\pi_1,\pi_2))=f(x(t),(u_1,u_2))$, if $(\pi_1,\pi_2)$
is a pattern of length $k=1$ of the form $(u_1,u_2)\in U$,
\item $f(x(t),(\pi_1,\pi_2))=f(f(x(t),(\pi'_1,\pi'_2)),(u_1,u_2))$, if  $(\pi_1,\pi_2)$ is a pattern of length $k\geq 2$ of the form
$(u_1\cdot\pi'_1,u_2\cdot\pi'_2)$ with
$(u_1,u_2)\in U$ and $(\pi'_1,\pi'_2)\in \Pi^{k-1}$.
\end{enumerate}
One defines $(f(x,\pi))_1\in \mathbb{R}^{n_1}$ and 
$(f(x,\pi))_2\in \mathbb{R}^{n_2}$ to be
the 1st and 2nd components of $f(x,\pi)\in \mathbb{R}^{n_1}\times\mathbb{R}^{n_2}=\mathbb{R}^{n}$, i.e:
$f(x,\pi)=((f(x,\pi))_1,f(x,\pi)_2)$.\\

In the following, we suppose that $K\in\mathbb{N}$
is an upper bound of the length of patterns.
The value of $K$ can be seen as a maximum number of time steps,
for which we compute the future behavior of the system (``horizon'').
%
%
We denote by $\Pi_1^{\leq K}$ 
(resp. $\Pi_2^{\leq K}$)
the expression $\bigcup_{1\leq k \leq K}\Pi_1^k$ 
(resp. $\bigcup_{1\leq k \leq K} \Pi_2^k$).
Likewise, we denote by $\Pi^{\leq K}$
the expression $\bigcup_{1\leq k \leq K} \Pi^k$.

\section{Control Synthesis Using Tiling}\label{sec:tiling}
\subsection{Tiling}\label{ss:cent0}
Let $R=R_1\times R_2$
be a rectangle.
We say that $\cal  R$
is a {\em (finite rectangular) tiling} of
$R$ if $\cal  R$ is of the form
$\{r_{i_1,i_2}\}_{i_1\in I_1, i_2\in I_2}$,
where 
$I_1$ and $I_2$ are given finite sets of positive integers, 
each $r_{i_1,i_2}$ is a sub-rectangle of $R$ of the form
$r_{i_1}\times r_{i_2}$, and $r_{i_1}, r_{i_2}$ 
are closed sub-intervals of $R_1$
and $R_2$ respectively. Besides, we have
$\bigcup_{i_1\in I_1}r_{i_1}=R_1$ and
$\bigcup_{i_2\in I_2}r_{i_2}=R_2$
(Hence $R=\bigcup_{i_1\in I_1, i_2\in I_2}r_{i_1,i_2}$).

We will refer to $r_{i_1}, r_{i_2}$ and $r_{i_1,i_2}$
as ``tiles'' of $R_1$, $R_2$ and $R$ respectively.
The same notions hold for rectangle $S$.


In the centralized context, given a rectangle $R$,
the {\em macro-step (backward reachability) control synthesis problem
with horizon $K$}
consists in finding a rectangle $S$
and a tiling ${\cal  S}=\{s_{i_1,i_2}\}_{i_1\in I_1, i_2\in I_2}$ of~$S$
such that, for each $(i_1,i_2)\in I_1\times I_2$,
there exists $\pi\in \Pi^{\leq K}$ such that:
$$f(s_{i_1,i_2},\pi)\subseteq R$$
(i.e., 
for all $x\in s_{i_1,i_2}$:
$f(x,\pi)\in R$).\\

\begin{figure}[!h]
  \centering
 \includegraphics[trim = 0cm 2.2cm 0cm 2cm, clip, scale=0.27]{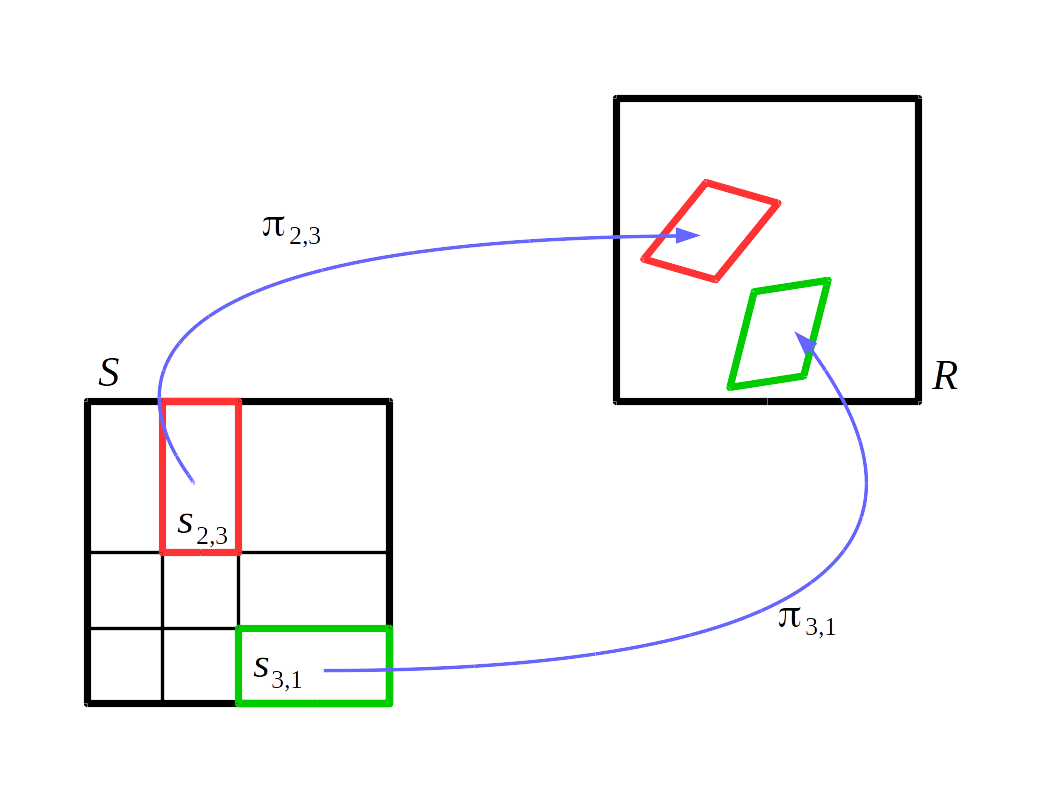}
  \caption{Mapping of tile $s_{2,3}$ to $R$ via pattern $\pi_{2,3}$,
and mapping of tile $s_{3,1}$ via  $\pi_{3,1}$.}
 \label{fig:reachability}
\end{figure}

This is illustrated in Figure~\ref{fig:reachability}.

\subsection{Parametric extension of tiling}\label{ss:extension}

In the following, we assume that the set $S$ we are looking for
is a {\em parametric extension} of $R$, denoted by $R+(a,a)$,
which is defined in the following.

Suppose that $R=R_1\times R_2$ is given
as well as a tiling
${\cal R}={\cal R}_1\times {\cal R}_2=\{r_{i_1}\times r_{i_2}\}_{i_1\in I_1, i_2\in I_2}
=\{r_{i_1,i_2}\}_{i_1\in I_1, i_2\in I_2}$.
$R_1$ can be seen as a product of $n_1$ closed intervals
of the form $[\ell,m]$. Consider a non negative real parameter $a$.
Let $(R_1+a)$ denote the corresponding
product of $n_1$ intervals of the form
$[\ell-a,m+a]$.\footnote{Actually, we will consider in the examples
that $(R_1+a)$ is a product of intervals of
the form $[\ell-a,m]$ where the interval is extended only
at its {\em lower} end, but the method is strictly identical.}
We define $(R_2+a)$ similarly.
Finally, we define $R+(a,a)$ as $(R_1+a)\times (R_2+a)$.

We now consider that $S$ is a (parametric) superset of $R$
of the form $R+(a,a)$.
We define a tiling ${\cal S}={\cal S}_1\times{\cal S}_2$
of $S$ of the form $\{s_{i_1}\times s_{i_2}\}_{i_1\in I_1,i_2\in I_2}$,
which is obtained from
${\cal R}={\cal R}_1\times{\cal R}_2=
\{r_{i_1}\times r_{i_2}\}_{i_1\in I_1,i_2\in I_2}$ by a simple extension,
as follows:

A tile $r_{i_1}$ (resp.~$r_{i_2}$) of ${\cal R}_1$ (resp. ${\cal R}_2$)
in ``contact''
with $\partial R_1$ (resp. $\partial R_2$)
is prolonged as a tile~$s_{i_1}$ (resp.~$s_{i_2}$)
in order to be in contact with $\partial (R_1+a)$
(resp. $\partial (R_2+a)$); 
a tile 
``interior'' to $R_1$ (i.e., with no contact with $\partial R_1$)
is kept unchanged, and coincides with $s_{i_1}$,
and similarly for $R_2$. 

\begin{figure}[!h]
  \centering
 \includegraphics[trim = 0cm 0.3cm 0cm 0.3cm, clip, scale=0.21]{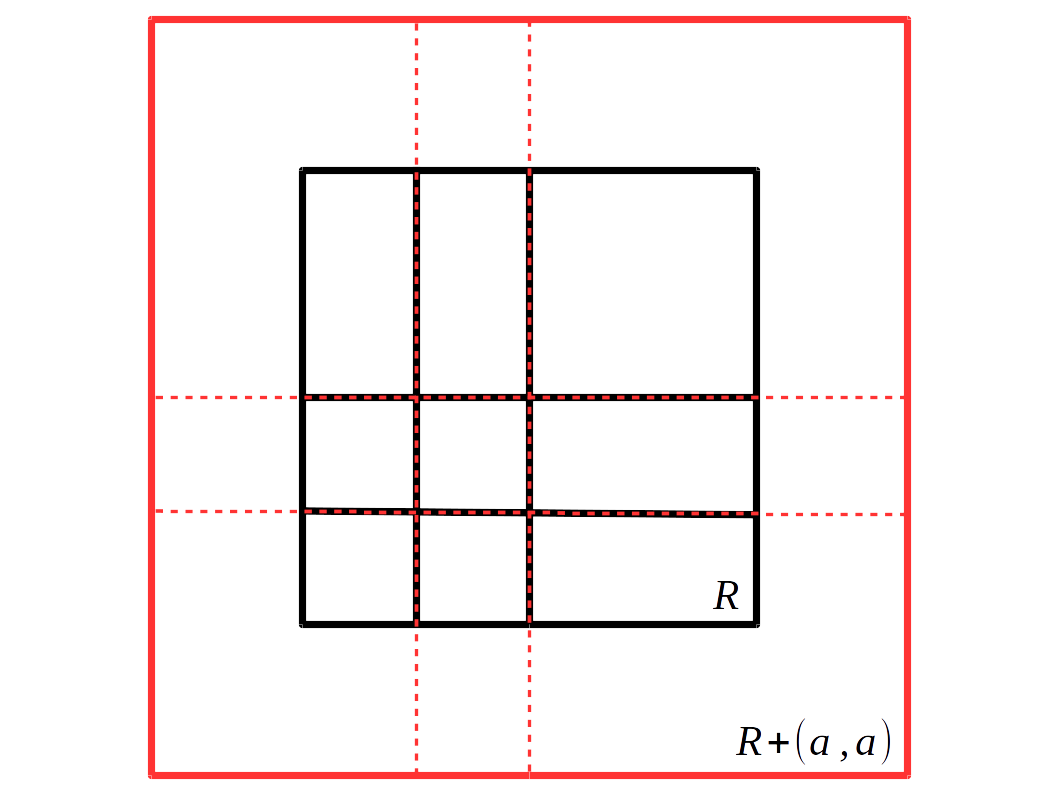}
  \caption{Tiling of $R+(a,a)$ induced by tiling ${\cal R}$ of $R$.}
 \label{fig:tiling}
\end{figure}

We denote the resulting tiling ${\cal S}$ by ${\cal R}+(a,a)$.
We also denote $s_{i_1}$ (resp.~$s_{i_2}$)
as $r_{i_1}+a$ (resp.~$r_{i_2}+a$) even if 
$r_{i_1}$ (resp.~$r_{i_2}$) is ``interior'' to $R_1$
(resp. $R_2$).
Likewise, we will denote $s_{i,j}$ as $r_{i,j}+(a,a)$.
Note that a tiling of $R$ of index set $I_1\times I_2$
induces a tiling of $R+(a,a)$ with the same index set
$I_1\times I_2$, hence the same number of tiles as $R$, for any $a\geq 0$.
This is illustrated in Figure~\ref{fig:tiling},
where the tiling of $R$ is represented with black continuous lines,
and the extended tiling of $R+(a,a)$ with red dashed lines.\\

\subsection{Generate and test tilings}\label{ss:gen}
By replacing $S$ with $R+(a,a)$ in the notions defined in
Section \ref{ss:cent0} 
the problem of macro-step control synthesis can now be reformulated as:\\

Finding a tiling ${\cal R}$ of $R$ 
which induces a
macro-step 
control of $R+(a,a)$  towards $R$,  for some $a\geq 0$; besides,
if we find such ${\cal R}$,
we want to compute the {\em maximum} value of $a$
for which the induced control exists.\\

This problem can be solved by
a simple ``generate and test'' procedure: one {\em generates} a
candidate tiling, then 
one {\em tests} if it satisfies the control property
(the control test procedure is explained  in Section 
\ref{ss:macro_cent}); if the test fails, one generate
another candidate, and so on iteratively.

In practice, the generation of a candidate ${\cal R}$
is done, starting from the trivial tiling (made of one tile equal to $R$),
then using successive {\em bisections} of $R$ 
until, either the control test succeeds (``success''), 
or the depth of bisection of the 
new candidate is greater than a given upper bound $D$
(``failure'').
See details in Appendix \ref{ss:refinement}.\\

{\em Remark 1.}
Note that, if the generate-and-test process stops with ``success'' for a tiling
${\cal R}$, then the tiling
${\cal R}_{D,uniform}$ also solves the problem,
where ${\cal R}_{D,uniform}$
is the ``finest'' tiling obtained 
by bisecting $D$ times all the $n$ components of $R$.
Since ${\cal R}_{D,uniform}$ has exactly $2^{nD}$ tiles, 
it is in general impractical to perform directly the control test on it.
From a theoretical point of view however, 
it is convenient to suppose that ${\cal R}={\cal R}_{D,uniform}$ 
for reducing the {\em worst case time complexity }
of the control synthesis procedure to the complexity 
of the control test part only
(see Section \ref{ss:macro_cent}).

\section{Centralized control}\label{sec:one-step}

\subsection{Tiling test procedure}\label{ss:macro_cent}

As seen in Section \ref{ss:extension},
the {\em (macro-step) 
control synthesis problem with horizon~K} 
consists in finding 
(the maximum value of) $a\geq 0$, 
and a tiling ${\cal R}=\{r_{i_1,i_2}\}_{i_1\in I_1, i_2\in I_2}$ of $R$
such that,
for each $(i_1,i_2)\in I_1\times I_2$ ,
there exists some $\pi\in \Pi^{\leq K}$ 
with
$$f(r_{i_1,i_2}+(a,a),\pi)\subseteq R.$$
In order to {\em test} if a tiling 
candidate ${\cal R}=\{r_{i_1,i_2}\}_{i_1\in I_1,i_2\in I_2}$  of $R$ satisfies the desired property,
we define, for each $(i_1,i_2)\in I_1\times I_2$:\\ 

$\Pi_{i_1,i_2}^{\leq K}=\{\pi\in \Pi^{\leq K}\ |\ f(r_{i_1,i_2},\pi)\subseteq R\}$.\\

When $\Pi_{i_1,i_2}^{\leq K}\neq\emptyset$, we define:

$a_{i_1,i_2}=\max_{\pi\in\Pi_{i_1,i_2}^{\leq K}}\max\{a\geq 0\ |\ f(r_{i_1,i_2}+(a,a),\pi)\ \subseteq R\}$

$\pi_{i_1,i_2}=argmax_{\pi\in\Pi_{i_1,i_2}^{\leq K}}\max\{a\geq 0\ |\ f(r_{i_1,i_2}+(a,a),\pi)\
\subseteq R\}$

$A=\min_{(i_1,i_2)\in I_1\times I_2}\{a_{i_1,i_2}\}$.\\

For each tile $r_{i_1,i_2}$ of $R$ and each $\pi\in\Pi^{\leq K}$, the test 
of inclusion $f(r_{i_1,i_2},\pi)\subseteq R$ can be done in time polynomial in $n$
when $f$ is affine.
Hence the test $\Pi_{i_1,i_2}^{\leq K}\neq\emptyset$
can be done in $O(N^K \cdot n^{\alpha})$  
since $\Pi^{\leq K}$ contains $O(N^K)$ elements.
The computation of $\max\{a\geq 0\ | f(r_{i_1,i_2}+(a,a),\pi)\subseteq R\}$
can be done by {\em linear programming} in time polynomial in $n$, 
the dimension of the state space.
The computation time of $\{a_{i_1,i_2}\}_{i_1\in I,i_2\in I_2}$, $\pi_{i_1,i_2}$,
and $A$ is thus in $O(N^K \cdot 2^{nD})$, where $D$ is the maximal 
depth of bisection.
Hence the complexity of testing
a candidate tiling ${\cal R}$ is in $O(N^K \cdot 2^{nD})$.
By Remark 1 above, the complexity of the 
control synthesis by generate-and-test is also in 
$O(N^K \cdot 2^{nD})$.

We have:
\begin{proposition}
%
Suppose that there exists a tiling 
${\cal R}=\{r_{i_1,i_2}\}_{i_1\in I_1, i_2\in I_2}$ of $R$ such that:
$$\forall (i_1,i_2)\in I_1\times I_2\ \ \Pi_{i_1,i_2}^{\leq K}\neq \emptyset.$$
Then ${\cal R}$ induces
a macro-step control
of horizon $K$ of $R+(A,A)$ towards $R$ with:
%
$$\forall (i_1,i_2)\in I_1\times I_2:\ 
\ \ f(r_{i_1,i_2}+(A,A),\pi_{i_1,i_2})\subseteq  R$$
where $A$ and $\pi_{i_1,i_2}$ are defined as above.

\end{proposition}

Once a candidate tiling ${\cal R}$ satisfying the control test
property is found, the generate-and-test procedure ends with
{\em success} (see Section \ref{ss:gen}),
and a set $S=R+(a^{(1)},a^{(1)})$ with $a^{(1)}=A$
has been found.
One can then {\em iterate} the ``generate and test'' procedure
in order to construct
an increasing sequence of nested rectangles of the form 
$R+(a^{(1)},a^{(1)})$, $R+(a^{(1)}+a^{(2)},a^{(1)}+a^{(2)})$, $\dots$,
which can all be driven to $R$,
as explained in Appendix \ref{ss:iteration}.

\begin{example}
\label{ex:ex2}
Consider the specification of a two-rooms appartment given in Example
\ref{ex:spec}. Set $R=[18.5,22]\times[18.5,22]$.
Let $D=1$ (the depth of bisection is at most 1),
and $K=4$ (the maximum length of patterns is~4).
We look for a centralized controller which will steer
the rectangle $S=[18.5-a,22]\times [18.5-a,22]$
to~$R$ with $a$ as large as possible, and stay in~$R$ indefinitely.
Using our implementation, the computation of the control synthesis
takes 4.14s of CPU time.

The method iterates successfully 15 times the macro-step control synthesis procedure.
We find $S=R+(a,a)$ with $a=53.5$, i.e. $S=[-35,22]\times[-35,22]$.
This means that any element of $S$ can be driven to $R$ within 15 macro-steps
of length (at most) 4, i.e., within $15\times 4=60$
units of time. 
Since each unit of time is of duration
$\tau=5$s, any trajectory starting from $S$ reaches $R$ within $60\times 5=300$s.
Once the trajectory $x(t)$ is in $R$, it returns in $R$
every macro-step of length (at most) 4, i.e., every $4\times 5=20$s.

These results are consistent with the simulation
given in Figure \ref{fig:simu_centralized} for the time evolution
of $(T_1,T_2)$ starting from $(12,12)$.
Simulations of the control, starting from $(T_1,T_2)=(12,12)$,
$(T_1,T_2)=(12,19)$ and $(T_1,T_2)=(22,12)$
are also given in the state space plane
in Figure \ref{fig:simu_centralized}.
\begin{figure}[h]
  \centering
  \begin{tabular}{cc}
 \includegraphics[trim = 0cm 0.3cm 0cm 1cm, clip, width=0.48\textwidth]{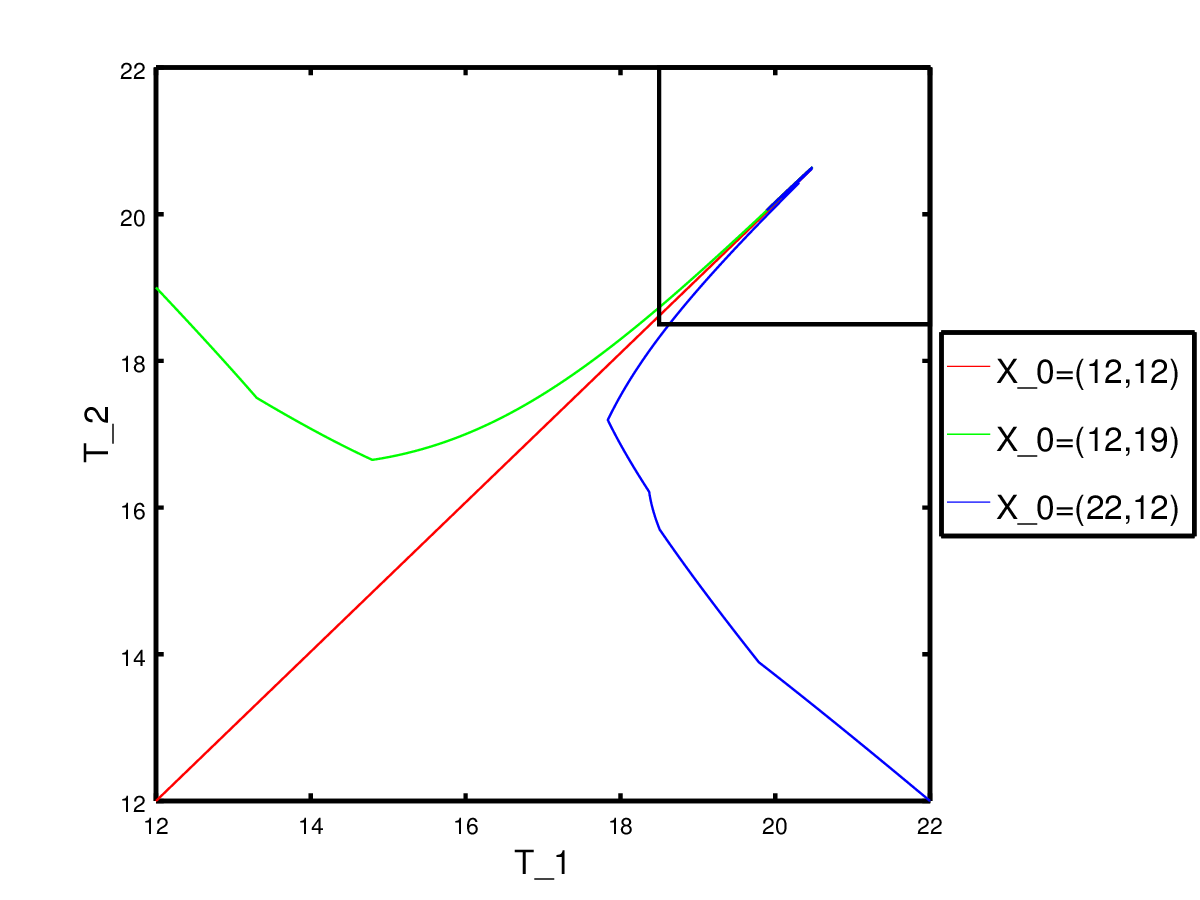} 
 &
  \includegraphics[trim = 0cm 0.3cm 0cm 1cm, clip, width=0.48\textwidth]{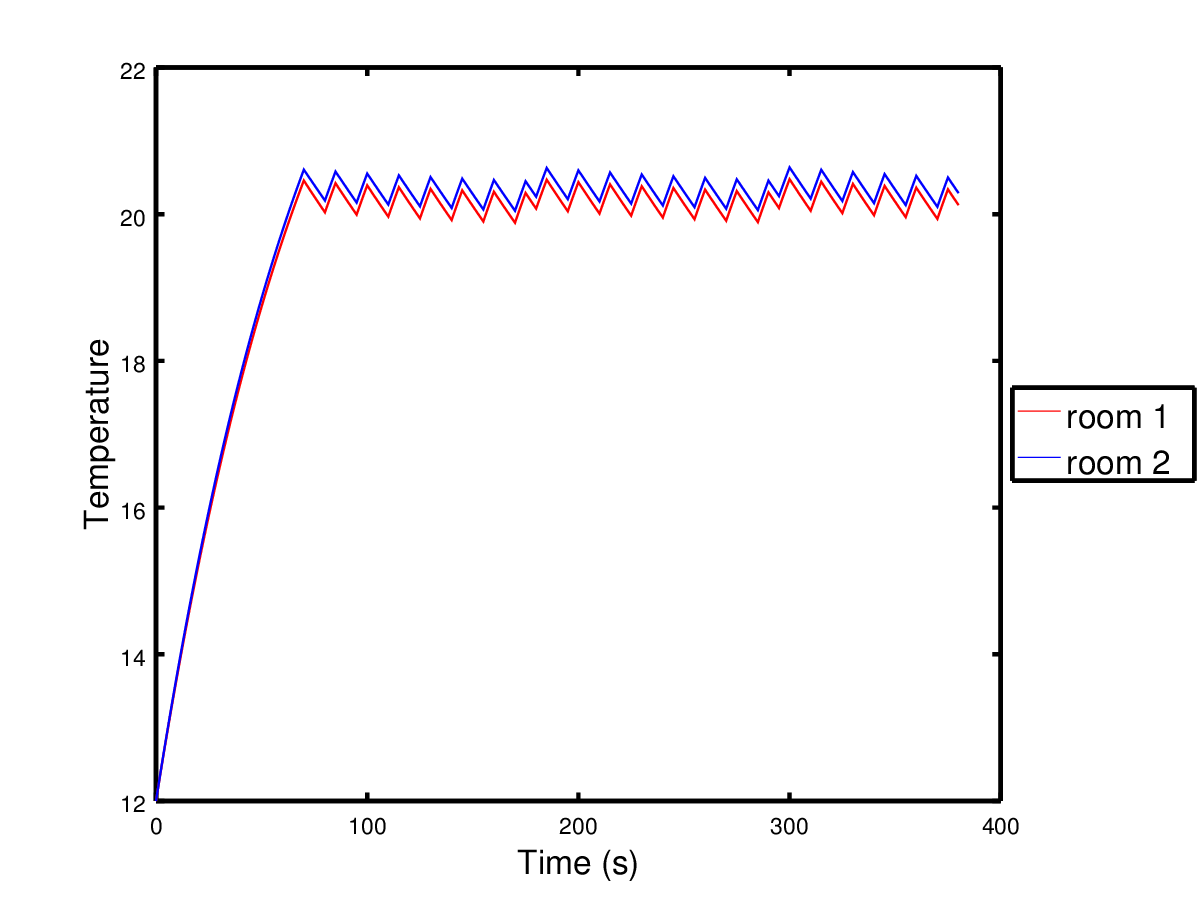}
  \end{tabular}

  \caption{Simulations of the centralized reachability controller for three different initial conditions plotted
  in the state space plane (left); simulation of the centralized reachability controller for the initial condition $(12,12)$ plotted 
   within time (right).}

  \label{fig:simu_centralized}
\end{figure}
%
%
%
 \end{example}
\subsection{Stability as a special case of reachability}\label{ss:special}
Instead of looking for a set of the form $S=R+(a,a)$ from
which $R$ is reachable via a macro-step,
let us consider the particular case where $S=R$ (i.e., $a=0$).

The problem is now to construct a tiling ${\cal R}=\{r_{i_1,i_2}\}_{i_1\in I_1, i_2\in I_2}$ of $R$
such that, for all $(i_1,i_2)\in I_1\times I_2$,
there exists a pattern $\pi_{i_1,i_2}\in \Pi^{\leq K}$ with $f(r_{i_1,i_2}, \pi_{i_1,i_2})\subseteq R$. 
If such a tiling ${\cal R}$ exists, then $x(t)\in R$ implies
$x(t+k)\in R$ for some $k\leq K$.\footnote{If $x(t)\in R$,
then $x(t)\in r_{i,j}$ for some $(i,j)\in I_1\times I_2$, 
hence $x(t+k)=f(x,\pi_{i,j})\in R$ for some $k\leq K$.}
Actually, we can slightly modify the procedure
in order to impose, additionally, that $\forall k\leq K\ x(t+k)\in R+\varepsilon$
for some $\varepsilon>0$ (see Section \ref{ss:macro_dist}).
It follows that $R+(\varepsilon,\varepsilon)$ is {\em stable} 
under the control induced by ${\cal R}$.

We can thus treat the stability control of $R$ 
as a special case  of reachability control.


\section{Distributed control}\label{sec:distr}
\subsection{Background}

In the distributed context,
given a set $R=R_1\times R_2$,
the {\em (macro-step) distributed 
control synthesis problem with horizon~K} 
consists in finding (the maximum value of) $a\geq 0$, and
a tiling ${\cal R}_1=\{r_{i_1}\}_{i_1\in I_1}$ of $R_1$
which induces a (macro-step) control
on $R_1+a$, a 
tiling ${\cal R}_2=\{r_{i_2}\}_{i_2\in I_2}$
which induces a (macro-step) control on $R_2+a$.

More precisely, we seek tilings ${\cal R}_1$ and ${\cal R}_2$ such that:
there exists $\ell\in{\mathbb N}$ such
that, for each $i_1\in I_1$ there exists a sequence
$\pi_1$ of $\ell$ modes in $U_1$,
and for each $i_2\in I_2$, 
a sequence $\pi_{2}$ 
of $\ell$ modes in $U_2$ such that:
$$(f((r_{i_1}+a)\times (R_2+a),(\pi_1,\pi_2)))_1\subseteq R_1 \ \wedge\ 
(f((R_1+a)\times (r_{i_2}+a),(\pi_1,\pi_2)))_2\subseteq R_2.$$

In order to synthesize a {\em distributed} strategy where
the control pattern $\pi_1$ is determined only by $i_1$
(regardless of the value of $i_2$), 
and the control pattern $\pi_2$ only by $i_2$
(regardless of the value of $i_1$),
we now define an {\em over-approximation} $X_{i_1}(a,\pi_1)$
for
$(f((r_{i_1}+a)\times(R_2+a),(\pi_1,\pi_2)))_1$,
and an {\em over-approximation} $X_{i_2}(a,\pi_2)$ for
$(f((R_1+a)\times (r_{i_2}+a),(\pi_1,\pi_2)))_2$.
The correctness of these over-approximations
relies on the existence of a 
fixed positive value for parameter $\varepsilon$.
Intuitively, $\varepsilon$ represents the width of
the additional margin (around $R+(a,a)$)
within which all the intermediate states lie when
a macro-step is applied to a point of $R+(a,a)$. \\

\subsection{Tiling test procedure}\label{ss:macro_dist}

Let $\pi_1^k$ (resp.$\pi_2^k$) denote the prefix of length $k$
of $\pi_1$ (resp.$\pi_2$), and $\pi_1(k)$ (resp. $\pi_2(k)$)
the $k$-th element of sequence $\pi_1$ (resp. $\pi_2$).
\begin{definition}
Consider an element $r_{i_1}$ (resp. $r_{i_2}$) of a tiling ${\cal R}_1$
(resp. ${\cal R}_2$) of $R_1$
(resp. $R_2$), and a sequence $\pi_1\in\Pi_1^{\leq K}$ (resp. $\pi_2\in\Pi_2^{\leq K}$)
of length $\ell_1$ (resp. $\ell_2$).
The {\em approximate 1st-component (resp. 2nd-component) sequence}
$\{X^k_{i_1}(a,\pi_1)\}_{0\leq k\leq \ell_1}$ (resp. $\{X^k_{i_2}(a,\pi_2)\}_{0\leq k\leq \ell_2}$)
is defined as follows:
\begin{itemize}
\item $X^0_{i_1}(a,\pi_1)=r_{i_1}+a$ and 
\item $X^{k}_{i_1}(a,\pi_1)=f_1(X^{k-1}_{i_1}(a,\pi_1),R_2+a+\varepsilon,\pi_1(k))$ for $1\leq k\leq \ell_1$
\end{itemize}
(resp. 
\begin{itemize}
\item $X^0_{i_2}(a,\pi_2)=r_{i_2}+a$ and 
\item $X^{k}_{i_2}(a,\pi_2)=f_2(R_1+a+\varepsilon,X^{k-1}_{i_2}(a,\pi_2),\pi_2(k))$
for $1\leq k\leq\ell_2$).
\end{itemize}
\end{definition}
We define the property
$Prop(a,i_1,\pi_1)$ of $\{X^k_{i_1}(a,\pi_1)\}_{0\leq k\leq \ell_1}$ by:
\begin{itemize}
\item  $X^{k}_{i_1}(a,\pi_1)\subseteq R_1+a+\varepsilon$ for $1\leq k\leq \ell_{1}-1$, and
\item  $X_{i_1}^{\ell_{1}}(a,\pi_1)\subseteq R_1$.
\end{itemize}
Likewise, we define the property
$Prop(a,i_2,\pi_2)$ of $\{X^k_{i_2}(a,\pi_2)\}_{0\leq k\leq \ell_2}$ by:
\begin{itemize}
\item  $X^{k}_{i_2}(a,\pi_2)\subseteq R_2+a+\varepsilon$ for $1\leq k\leq \ell_{2}-1$, and
\item  $X_{i_2}^{\ell_{2}}(a,\pi_2)\subseteq R_2$.
\end{itemize}

\begin{figure}[!h]
  \centering
  \begin{tabular}{cc}
 \includegraphics[trim = 2cm 1cm 2cm 2cm, clip, width=0.48\textwidth]{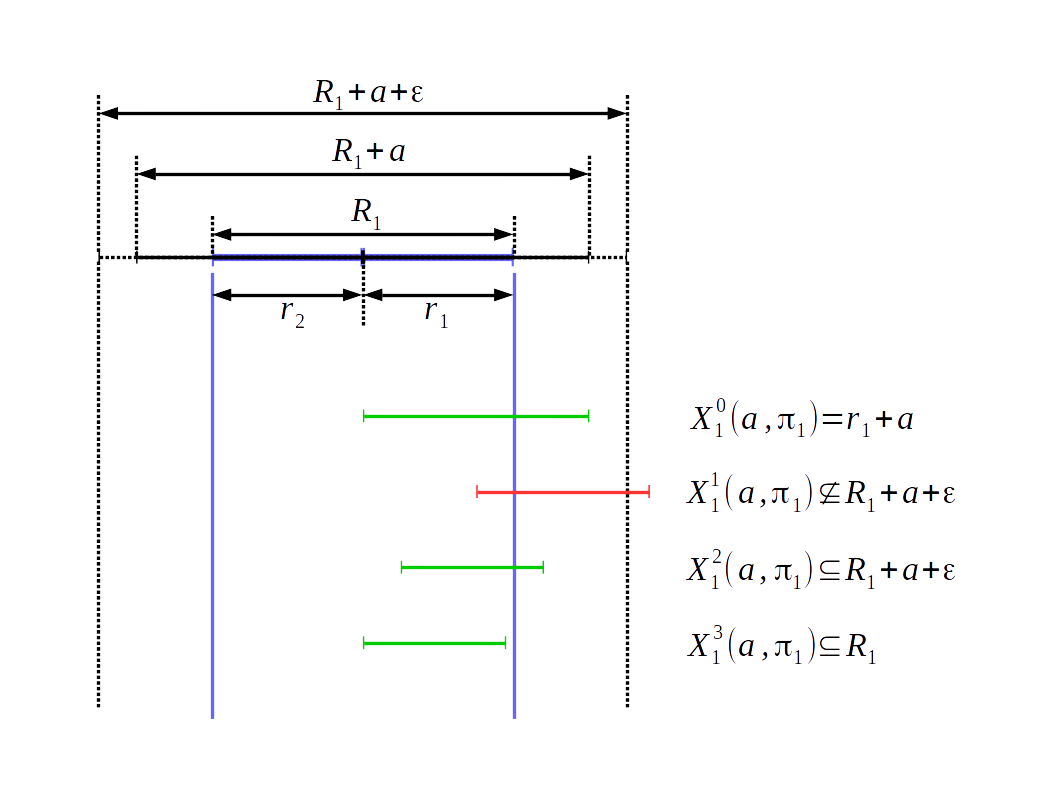}
 &
 \includegraphics[trim = 2cm 1cm 2cm 2cm, clip, width=0.48\textwidth]{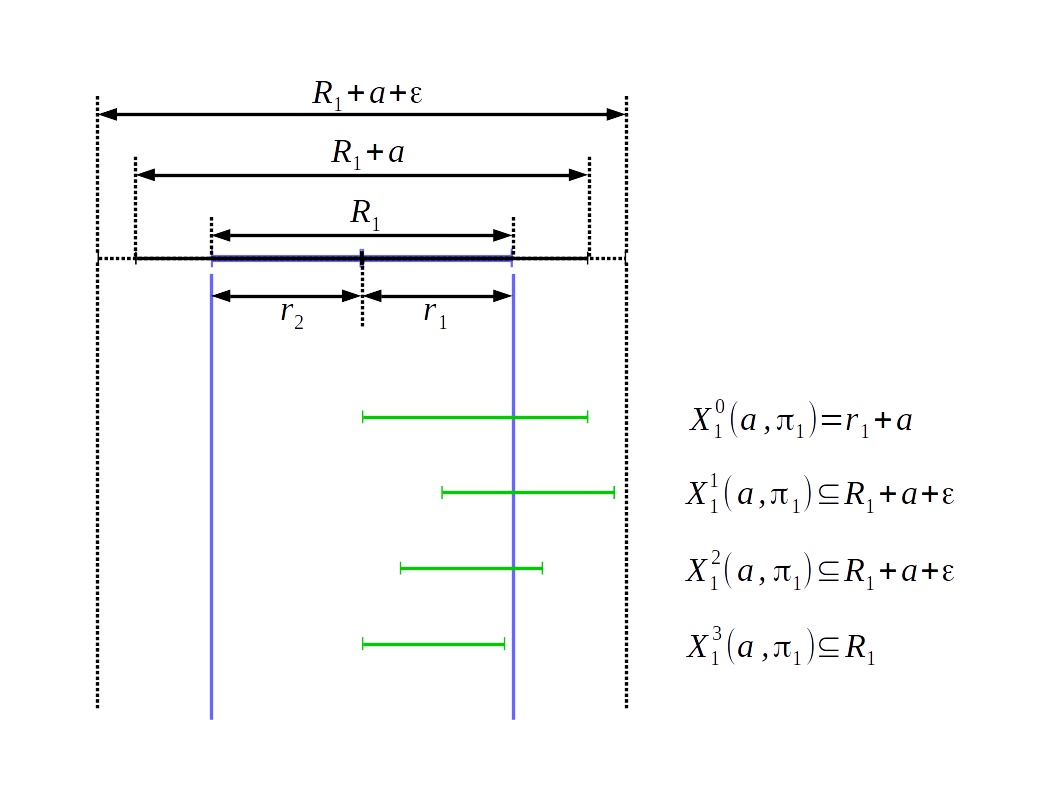}
  \end{tabular}

  \caption{Illustration of $\neg Prop(a,i_1,\pi_1)$ with $i_1=1$,
$|\pi_1|=\ell_1=3$
(left), and $Prop(a,i_1,\pi_1)$ (rigth).}
 \label{fig:Q1}
\end{figure}

Figure \ref{fig:Q1} illustrates property $Prop(a,i_1,\pi_1)$
for $\pi_1=(u_1\cdot v_1\cdot w_1)$,
$\ell_1=3$ and $i_1=1$:
in  the upper part, $Prop(a,i_1,\pi_1)$ is {\em not} satisfied because
$X_1^1(a,\pi_1)\subseteq R_1+a+\varepsilon$ is false
($X_1^2(a,\pi_1)\subseteq R_1+a+\varepsilon$ and
$X_1^3(a,\pi_1)\subseteq R_1$ are true);
in the lower part, $Prop(a,i_1,\pi_1)$
is satisfied.\\

Given a tiling ${\cal R}_1=\{r_{i_1}\}_{i_1\in I_1}$  of $R_1$,
we now define, for each $i_1\in I_1$, and each $k\in\{1,\dots,K\}$:\\

$\Pi_{i_1}^{k}=\{\pi_1\in \Pi_1^k\ |\ 
Prop(0,i_1,\pi_1)\}$.\\
\\
When $\Pi_{i_1}^k\neq\emptyset$, we define:

$a^k_{i_1}=\max_{\pi_1\in\Pi_{i_1}^{k}}\max\{a\geq 0\ |\ 
Prop(a,i_1,\pi_1)\}$

$\pi^k_{i_1}=argmax_{\pi_1\in\Pi_{i_1}^{k}}\max\{a\geq 0\ |\ 
Prop(a,i_1,\pi_1)\}$\\


Given ${\cal R}_2$, we define similarly: $\Pi_{i_2}^{k}$, $a^k_{i_2}$ and
$\pi^k_{i_2}$.\\

Suppose now, that: 

(H1) there exists $k_1\in \{1,\dots,K\}$ such that
$\forall i_1\in I_1: \Pi_{i_1}^{k_1}\neq \emptyset$.

(H2) there exists $k_2\in \{1,\dots,K\}$ such that
$\forall i_2\in I_2: \Pi_{i_2}^{k_2}\neq \emptyset$.\\

Then we define $a_1^{k_1}=\min_{i_1\in I_1}\{a_{i_1}^{k_1}\}$,
$a_2^{k_2}=\min_{i_2\in I_2}\{a_{i_2}^{k_2}\}$, and
$A=\min\{a_1^{k_1},a_2^{k_2}\}$.\\

NB1: Given a tiling ${\cal R}={\cal R}_1\times{\cal R}_2$,
(H1) means that the points of $R_1+A$ can be 
(macro-step) controlled to $R_1$ using patterns
which all have the {\em same length} $k_1$;
in other terms, all the macro-steps controlling $R_1+A$
contain the same number $k_1$ of elementary steps.
And symmetrically for (H2).\\

NB2: The determination of an appropriate value for $\varepsilon$
is for the moment done by hand, and is the result of a compromise:
if $\varepsilon$ is too small, then 
$f_1(r_{i_1}+a,R_2+a,u_1)\not\subseteq R_1+a+\varepsilon$; if
$\varepsilon$ is too big, $f_1(X_{i_1}^{k-1},R_2+a+\varepsilon,\pi_1(k))\not\subseteq R_1+a.$\\

Given a tiling ${\cal R}={\cal R}_1\times{\cal R}_2$ of $R$
and a real $\varepsilon>0$,
the problem of existence and computation of
$k_1$, $k_2$,
$\{\pi_{i_1}^{k_1}\}_{i_1\in I_1}$,
$\{\pi_{i_2}^{k_2}\}_{i_2\in I_2}$, and $A$
can be solved by
{\em linear programming} since $f_1$ and $f_2$ are affine.
Using the same kinds of calculation as in the centralized case (see
Section \ref{ss:macro_cent}), one can see that
the complexity of testing $\Pi_{i_1}^{k}\neq \emptyset$
and $\Pi_{i_2}^{k}\neq \emptyset$ for $1\leq k\leq K$,
checking (H1)-(H2),
generating $k_1$, $k_2$, $A$ and $\{\pi_{i_1}\}_{i_1\in I_1}$,
and $\{\pi_{i_2}\}_{i_2\in I_2}$
is in $O((\max(N_1,N_2))^K\cdot 2^{\max(n_1,n_2)D})$.
Hence the complexity of the control test procedure is also in
$O((\max(N_1,N_2))^K\cdot 2^{\max(n_1,n_2)D})$.\\

\begin{lemma}\label{prop:lemma}
Consider a tiling ${\cal R}={\cal R}_1\times{\cal R}_2$ of the
form $\{r_{i_1}\times r_{i_2}\}_{(i_1,i_2)\in I_1\times I_2}$.
Let $a\geq0$. We suppose that (H1) and (H2) hold,
and that, for all $i_1 \in I_1$,
$Prop(a,i_1,\pi_1)$ holds for some $\pi_1\in \Pi_1^{k_1}$,
and for all $i_2 \in I_2$, $Prop(a,i_2,\pi_2)$ holds for some $\pi_2\in \Pi_2^{k_2}$,
then we have:

\begin{itemize}
\item in case $k_1\leq k_2$:

$(f((r_{i_1}+a, R_2+a),(\pi_1^k,\pi_2^k)))_1\subseteq X_{i_1}^k(a,\pi_1)\subseteq R_1+a+\varepsilon$ and

$(f((R_1+a,r_{i_2}+a),(\pi_1^k,\pi_2^k)))_2\subseteq X_{i_2}^k(a,\pi_2)\subseteq R_2+a+\varepsilon$, 

\hspace*{\fill}for all $1\leq k\leq k_1$, and

$(f((r_{i_1}+a, R_2+a),(\pi_1^{k_1},\pi_2^{k_1})))_1\subseteq X_{i_1}^{k_1}(a,\pi_1)\subseteq R_1$,

\item in case $k_2 \leq k_1$:

$(f((r_{i_1}+a, R_2+a),(\pi_1^k,\pi_2^k)))_1\subseteq X_{i_1}^k(a,\pi_1)\subseteq R_1+a+\varepsilon$ and

$(f((R_1+a,r_{i_2}+a),(\pi_1^k,\pi_2^k)))_2\subseteq X_{i_2}^k(a,\pi_2)\subseteq R_2+a+\varepsilon$, 

\hspace*{\fill}for all $1\leq k\leq k_2$, and

$(f((R_1+a,r_{i_2}+a),(\pi_1^{k_2},\pi_2^{k_2})))_2\subseteq X_{i_2}^{k_2}(a,\pi_2)\subseteq R_2$.

\end{itemize}

\end{lemma}
The proof of given in Appendix \ref{sec:proof}.

At $t=0$, consider a point $x(0)=(x_1(0),x_2(0))$ of $R+(A,A)$,
and let us apply concurrently the strategy induced by ${\cal R}_1$
on $x_1$, and ${\cal R}_2$ on $x_2$.
After $k_1$ steps, by Lemma \ref{prop:lemma}, we obtain a point 
$x(k_1)=(x_1(k_1),x_2(k_1))\in R_1\times (R_2+A+\varepsilon)$.
Then, after $k_1$ steps,
we obtain again a point 
$x(2k_1)\in R_1\times (R_2+A+\varepsilon)$,
and so on iteratively.
Likewise, we obtain
points $x(k_2), x(2k_2),\dots$ which all belong to
$(R_1+A+\varepsilon)\times R_2$.
It follows that, after $\ell=lcm(k_1,k_2)$ steps, we obtain
a point $x(\ell)$ which belongs to $R_1\times R_2=R$.

\begin{theorem}
Suppose that there is a tiling ${\cal R}_1=\{r_{i_1}\}_{i_1\in I_1}$ of $R_1$,
a tiling ${\cal R}_2=\{r_{i_2}\}_{i_2\in I_2}$ of $R_2$,
and a positive real $\varepsilon$
such that
(H1) and (H2) hold, and let $k_1,k_2, A$ be defined as above.
Let $\ell=lcm(k_1,k_2)$ with $\ell=\alpha_1 k_1=\alpha_2 k_2$
for some $\alpha_1,\alpha_2\in \mathbb{N}$.

Then
${\cal R}_1$ induces a 
sequence of $\alpha_1$ macro-steps on $R_1+A$, and ${\cal R}_2$
a sequence of $\alpha_2$ macro-steps on $R_2+A$, such that, 
applied concurrently, we have,
for all $i_1\in I_1$ and $i_2\in I_2$:
$$(f((r_{i_1}+A)\times (R_2+A),\pi))_1\subseteq R_1 \ \wedge\ 
(f((R_1+A)\times (r_{i_2}+A),\pi))_2\subseteq R_2,$$
for some $\pi=(\pi_1,\pi_2)\in \Pi^{\ell}$ where $\pi_1$ (resp. $\pi_2$)
is of the form $\pi_1^1\cdots \pi_1^{\alpha_1}$
(resp. $\pi_2^1\cdots \pi_2^{\alpha_2}$)
with $\pi_1^i\in \Pi_1^{k_1}$  for all $1\leq i\leq \alpha_1$
(resp. $\pi_2^i\in \Pi_2^{k_2}$  for all $1\leq i\leq \alpha_2$).
Besides, for all prefix $\pi'$ of $\pi$, we have 
$$(f((r_{i_1}+A)\times (R_2+A),\pi'))_1\subseteq R_1+A+\varepsilon \ \wedge\ 
(f((R_1+A)\times (r_{i_2}+A),\pi'))_2\subseteq R_2+A+\varepsilon.$$
%

\end{theorem}

If (H1)-(H2) hold, there exists a control that
steers $R+(A,A)$ to $R$ in $\ell$ steps.
Letting $R'=R+(A,A)$, it is then possible to iterate the process
on $R'$ and, in case of success, 
generate a rectangle $R''=R'+(A',A')$ 
from which $R'$ would be
reachable in $\ell'$ steps, for some $A'\geq 0$ and $\ell'\in\mathbb{N}$.
And so on, iteratively, one generates
an increasing  sequence of nested control rectangles,
as in Section \ref{ss:macro_cent}.

\begin{example}
Consider again the specification of a two-rooms appartment given in Example
\ref{ex:spec}.
We consider the distributed control synthesis problem where
the 1st (resp. 2nd) state component corresponds to 
the temperature of the 1st (resp. 2nd) room $T_1$ (resp. $T_2$),
and the 1st (resp. 2nd) control mode component corresponds to the heater $u_1$
(resp. $u_2$) of the the 1st (resp. 2nd) room.

Set $R=R_1\times R_2=[18.5,22]\times[18.5,22]$.
Let $D=3$ (the depth of bisection is at most 3),
and $K=10$ (the maximum length of patterns is 10).
The parameter $\varepsilon$ is set to value $1.5^\circ C$.
We look for a distributed controller 
which steers any temperature state
in $S=S_1\times S_2=[18.5-a,22]\times [18.5-a,22]$
to $R$ with $a$ as large as possible, 
then maintain it in~$R$ indefinitely.

Using our implementation, the computation of the control synthesis
takes 220s of CPU time.

The method iterates 8 times the macro-step control synthesis procedure.
We find $S=[18.5-a,22]\times [18.5-a,22]$ with $a=6.5$, i.e. 
$S=[12,22]\times[12,22]$.
This means that any element of $S$ can be driven to $R$ within 8 macro-steps
of length (at most) 10, i.e., within $8\times 10=80$
units of time. 
Since each unit of time is of duration
$\tau=5$s, any trajectory starting from $S$ reaches 
$R$ within $80\times 5=400$s.
The trajectory is then guaranteed to always stay (at each discrete time $t$)
in $R+(\varepsilon,\varepsilon)=[17,23.5]\times[17,23.5]$.

These results are consistent with the simulation given in
Figure \ref{fig:simu_distri} showing the time evolution
of $(T_1,T_2)$ starting from $(12,12)$.
Simulations of the control
are also given in the state space plane, in
Figure \ref{fig:simu_distri}, for initial
states $(T_1,T_2)=(12,12)$,
$(T_1,T_2)=(12,19)$ and $(T_1,T_2)=(22,12)$.

Not surprisingly, the performance guaranteed by the distributed approach ($a = 6.5$, 
attainability of $R$ in $400$s) are worse than those guaranteed by the centralized 
approach of Example \ref{ex:ex2} ($a=53.5$, attainability of $R$ in $300$s).
However, unexpectedly, the CPU computation time in the distributed approach ($220$s) is 
here worse than the CPU time of the centralized approach ($4.14$s). This relative 
inefficiency is due to the small size of the example.

\begin{figure}[!h]
  \centering
  \begin{tabular}{cc}
   \includegraphics[trim = 0cm 0.3cm 0cm 1cm, clip, width=0.48\textwidth]{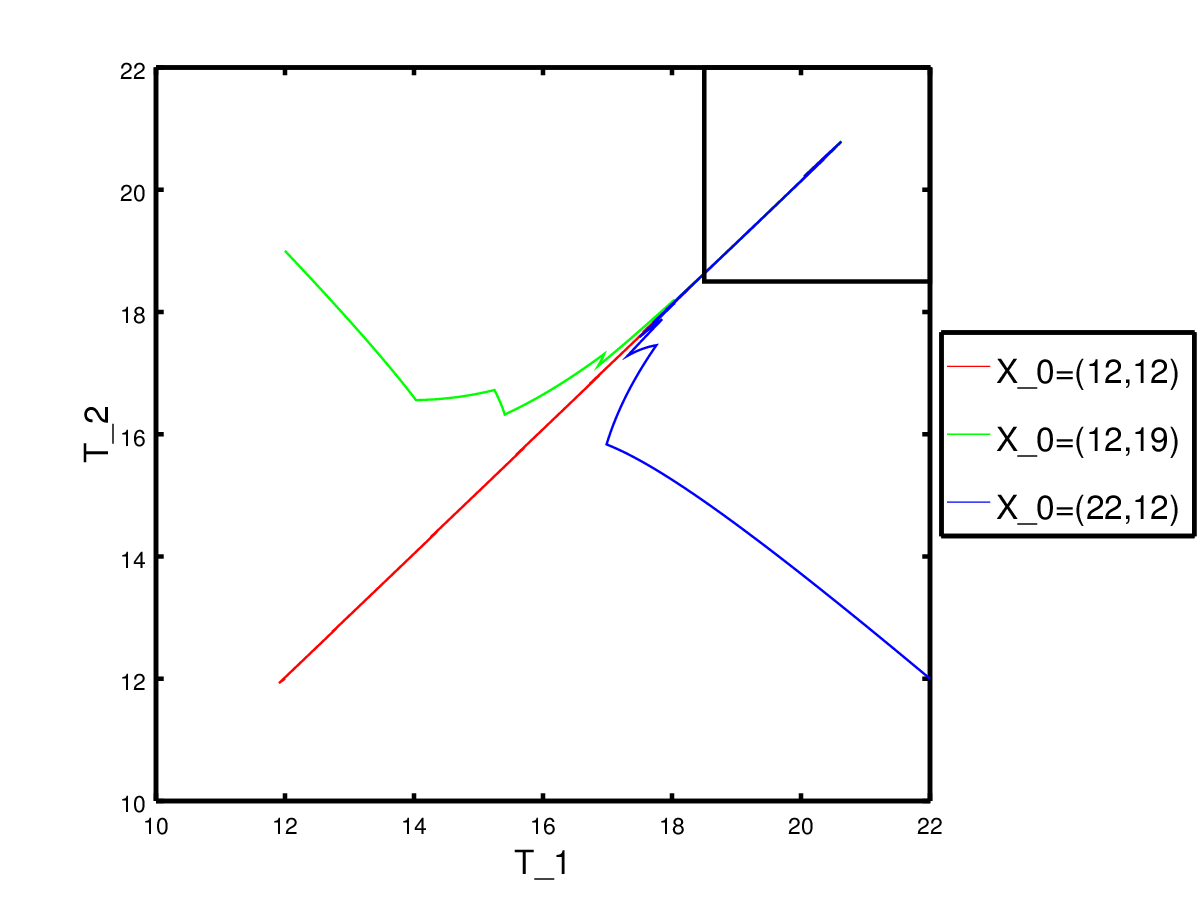}
   &
   \includegraphics[trim = 0cm 0.3cm 0cm 1cm, clip, width=0.48\textwidth]{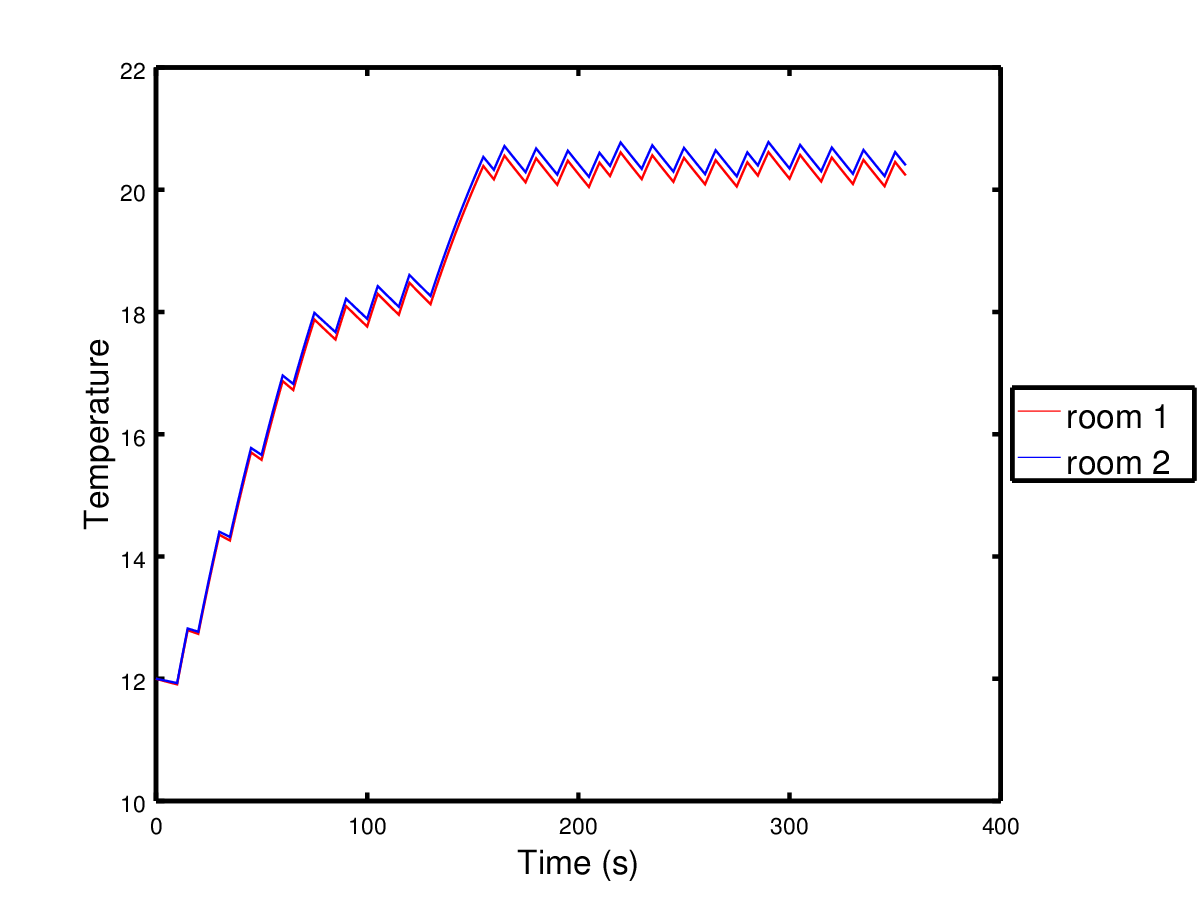}
  \end{tabular}

  \caption{Simulations of the distributed reachability controller for three different initial conditions plotted
  in the state space plane (left); simulation of the distributed reachability controller for the initial condition $(12,12)$ plotted 
  within time (right).}
 \label{fig:simu_distri}
\end{figure}

%
%
 \end{example}


\section{Case study}

This case study, proposed by the Danish company Seluxit, aims at controlling 
the temperature of an eleven rooms house, heated by geothermal energy.

The {\em continuous} dynamics of the system is the following:
\begin{equation}
 \frac{d}{dt} T_i(t) = \sum_{j=1}^n A_{i,j}^d (T_j(t) - T_i(t)) + B_i(T_{env} (t) - T_i(t)) + H_{i,j}^v.v_j 
\end{equation}

The temperatures of the rooms are the $T_i$.
The matrix $A^d$ contains the heat transfer coefficients between the rooms, matrix $B$ contains
the heat transfer coefficients betweens the rooms and the external temperature, set to $T_{env} = 10^\circ C$ for the 
computations. The control matrix $H^v$ contains the effects of the control on the room temperatures, and the control
variable is here denoted by $v_j$. We have $v_j = 1$ (resp. $v_j = 0$) if the heater
in room $j$ is turned on (resp. turned off). We thus have $n=11$
and $N=2^{11} = 2048$ switching modes.

Note that the matrix $A^d$ is parametrized by the open of closed state of the doors 
in the house. In our case, the average between closed and open matrices was taken 
for the computations. The exact values of the coefficients are given in
\cite{larsen2015online}.
The controller has to select which heater to turn on in the eleven rooms. Due to a limitation
of the capacity supplied by the geothermal device, the $11$ heaters cannot be turned on at the same time. 
In our case, we set to $4$ the maximum number of heaters turned on at the same time. \\

We consider the distributed control synthesis problem where
the 1st (resp. 2nd) state component corresponds to 
the temperatures of rooms 1 to 5 (resp. 6 to 11),
and the 1st (resp. 2nd) control mode component corresponds to the heaters 
of rooms 1 to 5 (resp. 6 to 11).
Hence $n_1=5, n_2=6, N_1=2^5, N_2=2^6$. We impose
that at most $2$ heaters are switched on at the same time in the
$1^{st}$ sub-system, and at most $2$ in the $2^{nd}$ sub-system.

Let $D=1$ (the depth of bisection is at most 1),
and $K=4$ (the maximum length of patterns is 4).
The parameter $\varepsilon$ is set to value $0.5^\circ C$.
The sampling time is $\tau=15$ min.

We look for a distributed controller 
which steers any temperature state in
the rectangle $S=[18-a,22]^{11}$
to $R=[18,22]^{11}$ with $a$ as large as possible, 
then maintain the temperatures in~$R$ indefinitely.

Using our implementation, the computation of the control synthesis
takes around 20 hours of CPU time.

The method iterates 15 times successfully
the macro-step control synthesis procedure.
We find $S=[18-a,22]^{11}$ with $a=4.2$, i.e. 
$S=[13.8,22]^{11}$.
This means that any element of $S$ can be driven into $R$ within 15 macro-steps
of length (at most) 4, i.e., within $15\times 4=60$
units of time. 
Since each unit of time is of duration
$\tau=15$ min, any trajectory starting from $S$ attains 
$R$ within $60\times 15=900$ min.
The trajectory is then guaranteed to stay
in $R+(\varepsilon,\varepsilon)=[17.5,22.5]^{11}$.

\begin{figure}[!h]
  \centering
 \includegraphics[trim = 0cm 1cm 0cm 0cm, clip, scale = 0.4]{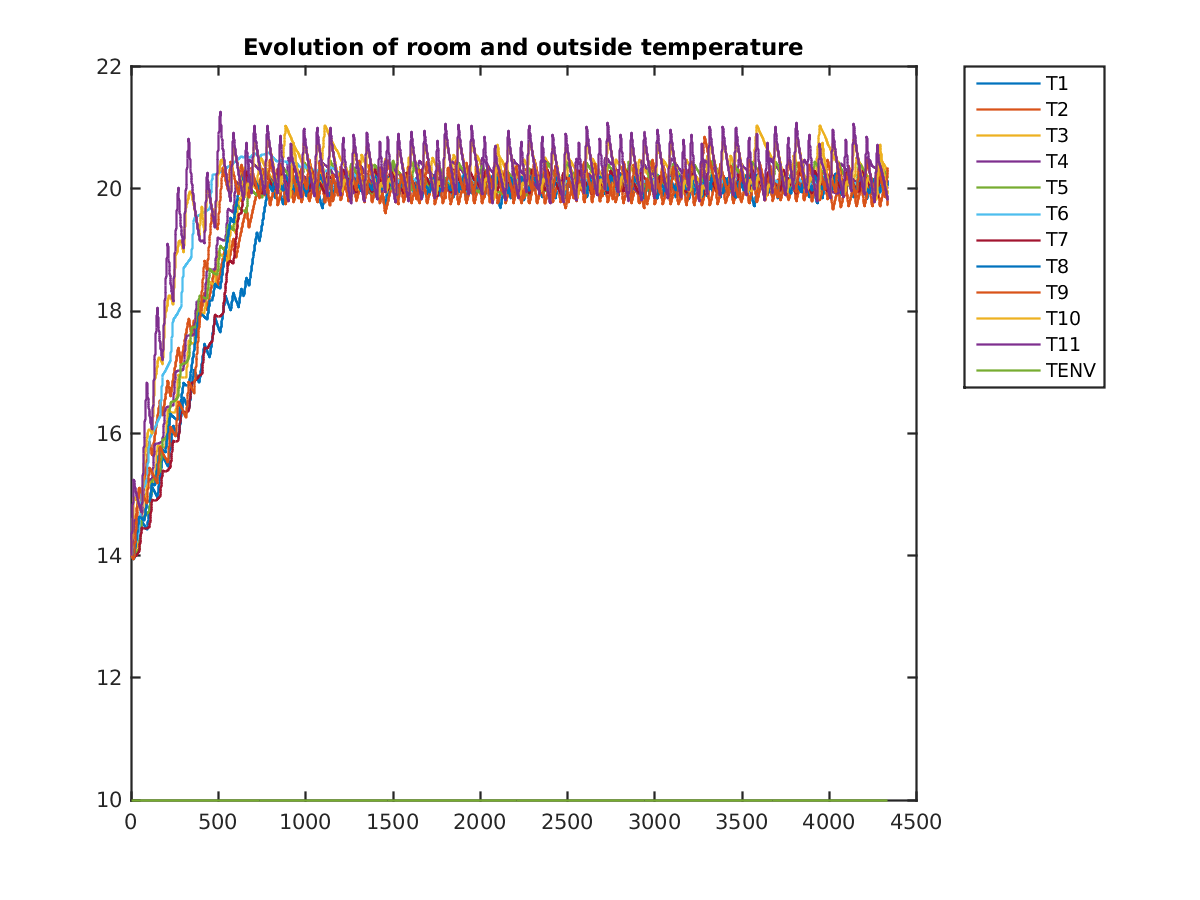}
 \caption{Simulation of the Seluxit case study 
plotted with time (in min) for $T_{env}=10^{\circ} C$.}
 \label{fig:reach_11rooms_10}
\end{figure}

These results are consistent with the simulation given in
Figure \ref{fig:reach_11rooms_10} showing the time evolution
of the temperature of the rooms, starting from $14^{11}$.

Robustness simulations for our controller
are given in Appendix \ref{sec:robustness}.

\section{Final Remarks}

In this paper, we have proposed a distributed approach 
for control synthesis and applied it to a real floor heating system.
To our knowledge,
this is the first time that attainability and stability properties  
are guaranteed for a case study of this size. The method can
be extended to take into account obstacles and safety constraints.
We are currently investigating an 
extension of the method to systems with non linear
dynamics and varying parameters.

\nocite{*}
\bibliographystyle{eptcs}
\bibliography{cassting}

\section{Appendix: Control Synthesis}
\subsection{Tiling refinement}\label{ss:refinement}
Let us now explain how we find a tiling ${\cal R}$ of $R$
such that $\Pi_{i_1,i_2}\neq \emptyset$.
We focus on the centralized case, but the distributed case is similar.
We start from the trivial tiling ${\cal R}^0=\{R\}$ which consists
of just one tile equal to $R$. If $f(R,\pi)\subseteq R$ for some $\pi\in \Pi^{\leq K}$,
then ${\cal R}^0$ is the desired tiling. Otherwise,
we refine ${\cal R}^0$ by {\em bisection}, which gives
a tiling ${\cal R}^1$ of the form $\{r_{(i,1),(j,2)}\}_{1\leq i,j\leq n}$.
If, for all $1\leq i,j\leq n$ there exists some $\pi\in \Pi^{\leq K}$ such that
$f(r_{(i,1),(j,2)},u)\subseteq R$, then ${\cal R}^1$ is the desired tiling.
Otherwise, there exist some ``bad'' tiles of the form $r_{(i,1),(j,2)}$
with $1\leq i,j\leq n$ such that $\forall \pi\in \Pi^{\leq K}\ f(r_{(i,1),(j,2)},\pi)\not\subseteq R$;
we then transform ${\cal R}^1$ into ${\cal R}^2$ by bisecting
all the bad tiles. And so on iteratively, 
we produce tilings ${\cal R}^1, {\cal R}^2,\cdots, {\cal R}^d$ until
either no bad tiles remain in ${\cal R}^d$ ({\em success}) 
or $d$ is greater than
the upper bound $D$ of depth of bisection ({\em failure}).

\subsection{Iterated macro-step control synthesis}\label{ss:iteration}
\begin{figure}[!h]
  \centering
 \includegraphics[trim = 0cm 1cm 0cm 1cm, clip, scale=0.25]{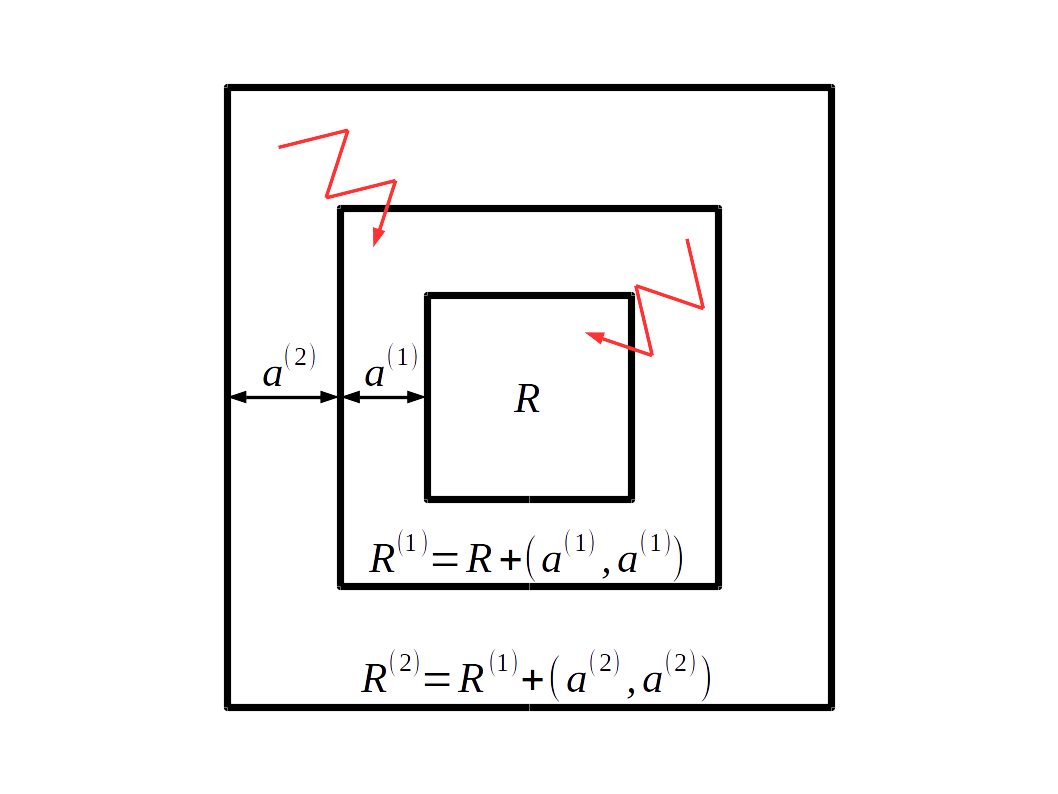}
  \caption{Iterated control of $R^{(1)}=R+(a^{(1)},a^{(1)})$ towards $R$,
and $R^{(2)}=R^{(1)}+(a^{(2)},a^{(2)})$ towards $R^{(1)}$.}
 \label{fig:iteration}
\end{figure}

Suppose that we are given an objective rectangle $R=R_1\times R_2$.
If the one-step control synthesis described 
in Section \ref{ss:refinement} succeeds,
then there is a positive null real $a^{(1)}=A$
and a tiling ${\cal R}$ of $R$ which
induces a control steering all the points of $R^{(1)}=R+(a^{(1)},a^{(1)})$
to~$R$ in one step.
Now the macro-step control synthesis can be reapplied to
$R^{(1)}$.
If it succeeds again, then it produces
a tiling ${\cal R}^{(1)}$ of $R^{(1)}$ which induces a control
that steers 
$R^{(2)}=R^{(1)}+(a^{(2)},a^{(2)})$ to $R^{(1)}$ 
for some $a^{(2)}\geq 0$.
And so on, the iterated application
of macro-step control synthesis outputs a sequence of 
tilings ${\cal R}^{(i)}$
which induce a control that steers 
$R^{(i+1)}=R+(\Sigma_{j=1}^{i+1}a^{(j)},\Sigma_{j=1}^{i+1}a^{(j)})$ 
to $R^{(i)}$, for some $a^{(j)}\geq 0$ ($1\leq j\leq i+1$).
We thus synthesize a control which steers $R^{(i+1)}$
to $R$ in at most $i+1$ macro-steps ($i\geq 0$),
using an increasing sequence of 
nested rectangles around $R$.
This is illustrated in Figure \ref{fig:iteration}, for $i=1$.

The iteration process halts
at step, say $m$, when the last
macro-step control synthesis fails because
the maximum bisection depth $D$ is reached while ``bad'' tiles still remain
(see Section \ref{ss:refinement}). We also stop
the process when the last macro-step control synthesis outputs
a real~$a^{(m)}$ which
is smaller than a given bound $\eta>0$: this is because the 
sequence of controllable rectangles around $R$ seems to
approach a limit.

\section{Appendix: Proof of Lemma \ref{prop:lemma}}\label{sec:proof}

Let us suppose $k_1 \leq k_2$.
Let us denote by $P_{i_1}^1(k)$ the property
\begin{equation*}
(f((r_{i_1}+a,R_2+a),(\pi_1^k,\pi_2^k)))_1 \subseteq X_{i_1}^k
 \label{eq:firstinc}
\end{equation*}
and by $P_{i_1}^2(k)$
\begin{equation*}
X_{i_1}^k \subseteq R_1 + a + \varepsilon
 \label{eq:secondinc}
\end{equation*}
and similarly for $P_{i_2}^1(k)$ and $P_{i_2}^2(k)$.

We are going to show by induction on $k$ the property $P(k)$:
\begin{equation*}
\forall i_1 \in I_1, \ P_{i_1}^1(k) \wedge  P_{i_1}^2(k) \quad \text{and} \quad \forall i_2 \in I_2, \ P_{i_2}^1(k) \wedge  P_{i_2}^2(k). 
\end{equation*}

Let us first consider the case $k = 1$.
Let us prove $\forall i_1 \in I_1, \ P_{i_1}^1(k) \wedge  P_{i_1}^2(k)$. (The proof is 
similar for $\forall i_2 \in I_2, \ P_{i_2}^1(k) \wedge  P_{i_2}^2(k)$.)
Let us show that $(f((r_{i_1}+a,R_2+a),(\pi_1^k,\pi_2^k)))_1 \subseteq X_{i_1}^k$
and $X_{i_1}^k \subseteq R_1 + a + \varepsilon$.

For $k = 1$,
$\pi_1^k$ and $\pi_2^k$ are of the form $u_1$ and $u_2$. We have:
\begin{enumerate}
 \item $(f((r_{i_1}+a,R_2+a),(\pi_1^k,\pi_2^k)))_1 = f_1(r_{i_1} + a, R_2 + a, u_1)$
 \item $X_{i_1}^1 = f_1(X_{i_1}^0,R_2 + a + \varepsilon, u_1) = f_1(r_{i_1} + a ,R_2 + a + \varepsilon, u_1) $
\end{enumerate}

Hence $(f((r_{i_1}+a,R_2+a),(\pi_1^k,\pi_2^k)))_1 \subseteq X_{i_1}^k$ holds
for $k=1$. And $X_{i_1}^k \subseteq R_1 + a + \varepsilon$ because 
of $Prop(a,i_1,\pi_1)$.

Let us now suppose that $k>1$ and that $P(k-1)$ holds. Let us prove $P(k)$.

Properties $P_{i_1}^2(k)$ and $P_{i_2}^2(k)$ are true for all $i_1,i_2$ because, 
by construction, the sequence $X_{i_1}^k$ (resp. $X_{i_2}^k$)
satisfies $Prop(a,i_1,\pi_1)$ (resp. $Prop(a,i_2,\pi_2)$).
Let us prove $P_{i_1}^1(k)$ and $P_{i_2}^1(k)$:

\begin{eqnarray*}
 (f(r_{i_1} +a,R_2+a,(\pi_1^k,\pi_2^k)))_1 & = & (f(f((r_{i_1}+a,R_2+a),(\pi_1^{k-1},\pi_2^{k-1})),(\pi_1{(k)},\pi_2{(k)})))_1 \\
  & = & f_1(\lbrack f((r_{i_1}+a,R_2+a),(\pi_1^{k-1},\pi_2^{k-1})) \rbrack_1, \\
 & &  \lbrack f((r_{i_1}+a,R_2+a),(\pi_1^{k-1},\pi_2^{k-1})) \rbrack_2,\pi_1{(k)} ). 
\end{eqnarray*} 
 
 Note that  the first argument of $f_1$ in the last expression satisfies
  $\lbrack f((r_{i_1}+a,R_2+a),(\pi_1^{k-1},\pi_2^{k-1}))\rbrack_1 \subseteq X_{i_1}^k $
  by $P_{i_1}^1(k-1)$. Besides, the second argument
  satisfies $\lbrack f((r_{i_1}+a,R_2+a),(\pi_1^{k-1},\pi_2^{k-1})) \rbrack_2
  \subseteq \bigcup_{j_2 \in I_2}X_{j_2}^{k-1} \subseteq R_2 + a + \varepsilon$, because 
  \begin{enumerate}
   \item $r_{i_1} + a \subseteq R_{1}+a$
   \item $\bigcup_{j_2 \in I_2} X_{j_2}^{k-1} \subseteq R_2 + a + \varepsilon$ since $X_{j_2}^{k-1} \subseteq R_2 + a + \varepsilon$ \quad (by $P_{j_2}^2(k-1)$ which holds for all $j_2$)
   \item $\lbrack f((R_1 + a , r_{j_2} + a), (\pi_1^{k-1}, \pi_2^{k-1})) \rbrack_2 
  \subseteq X_{j_2}^{k-1}$ \quad (by $P_{j_2}^1(k-1)$).
  \end{enumerate}


 Hence 
 \begin{eqnarray*}
  f_1(\lbrack f((r_{i_1}+a,R_2+a),(\pi_1^{k-1},\pi_2^{k-1})) \rbrack_1,\lbrack f((r_{i_1}+a,R_2+a),(\pi_1^{k-1},\pi_2^{k-1})) \rbrack_2,\pi_1^{(k)} ) \\
  \subseteq f_1(X_{i_1}^{k-1},R_2 + a + \varepsilon,\pi_1(k)) = X_{i_1}^k
 \end{eqnarray*}

 We have thus proved $P_{i_1}^1(k)$:
 \begin{equation*}
 (f(r_{i_1} +a,R_2+a,(\pi_1^k,\pi_2^k)))_1 \subseteq X_{i_1}^k
\end{equation*}

This completes the proof of $\forall i_1 \in I_1,\  P_{i_1}^1(k) \wedge P_{i_1}^2(k)$
We prove $\forall i_2 \in I_2,\  P_{i_2}^1(k) \wedge P_{i_2}^2(k)$
similarly, which achieves the proof of $P(k)$.

The proof of $(f((r_{i_1}+a,R_2+a),(\pi_1^{k_1},\pi_2^{k_1})))_1 \subseteq X_{i_1}^{k_1}(a,\pi_1) \subseteq R_1$
is similar.

%
%

\section{Appendix: Robustness Experiments}\label{sec:robustness}

\begin{figure}[h]
  \centering
 \includegraphics[trim = 0cm 1cm 0cm 0cm, clip, scale = 0.4]{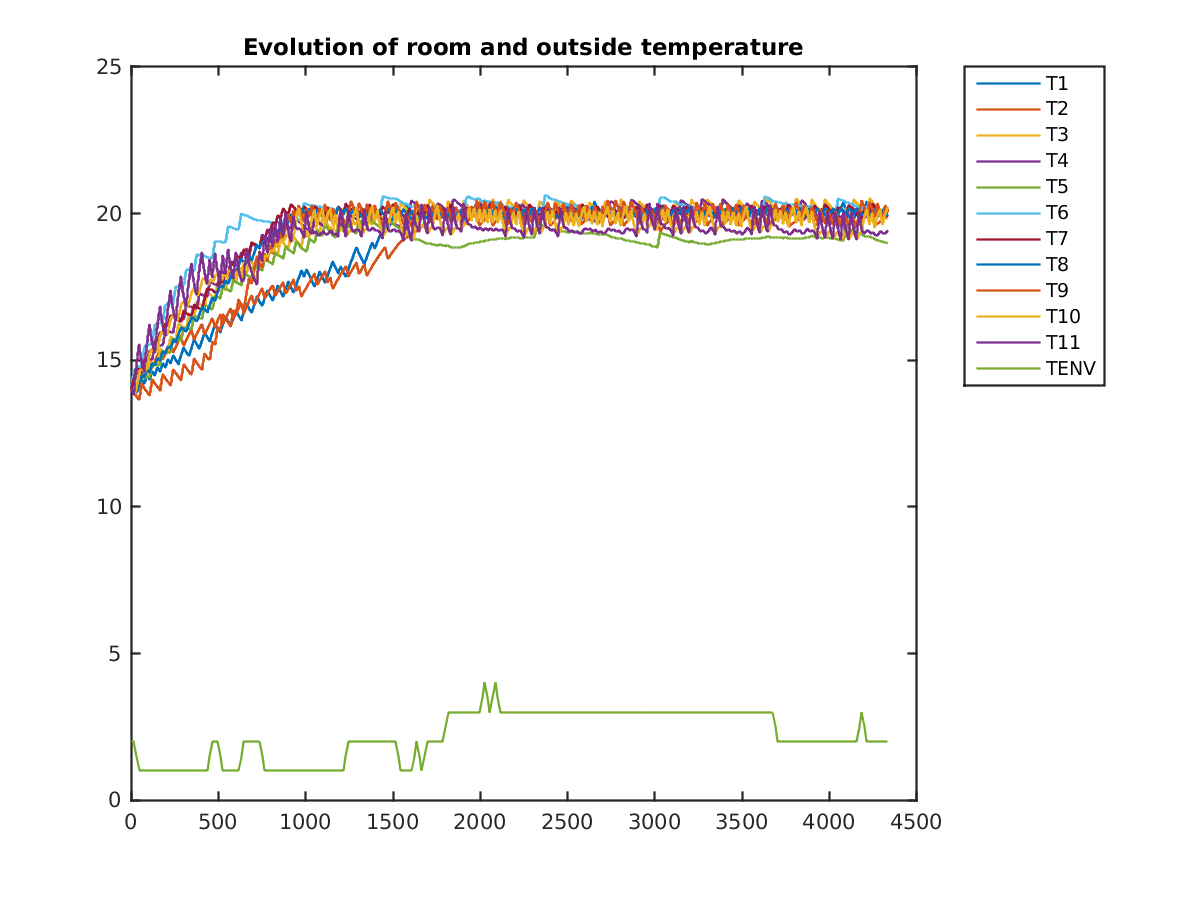}
 \caption{Simulation of the Seluxit case study in the soft winter scenario.}
 \label{fig:reach_11rooms}
\end{figure}

\begin{figure}[h]
  \centering
 \includegraphics[trim = 0cm 1cm 0cm 0cm, clip, scale = 0.4]{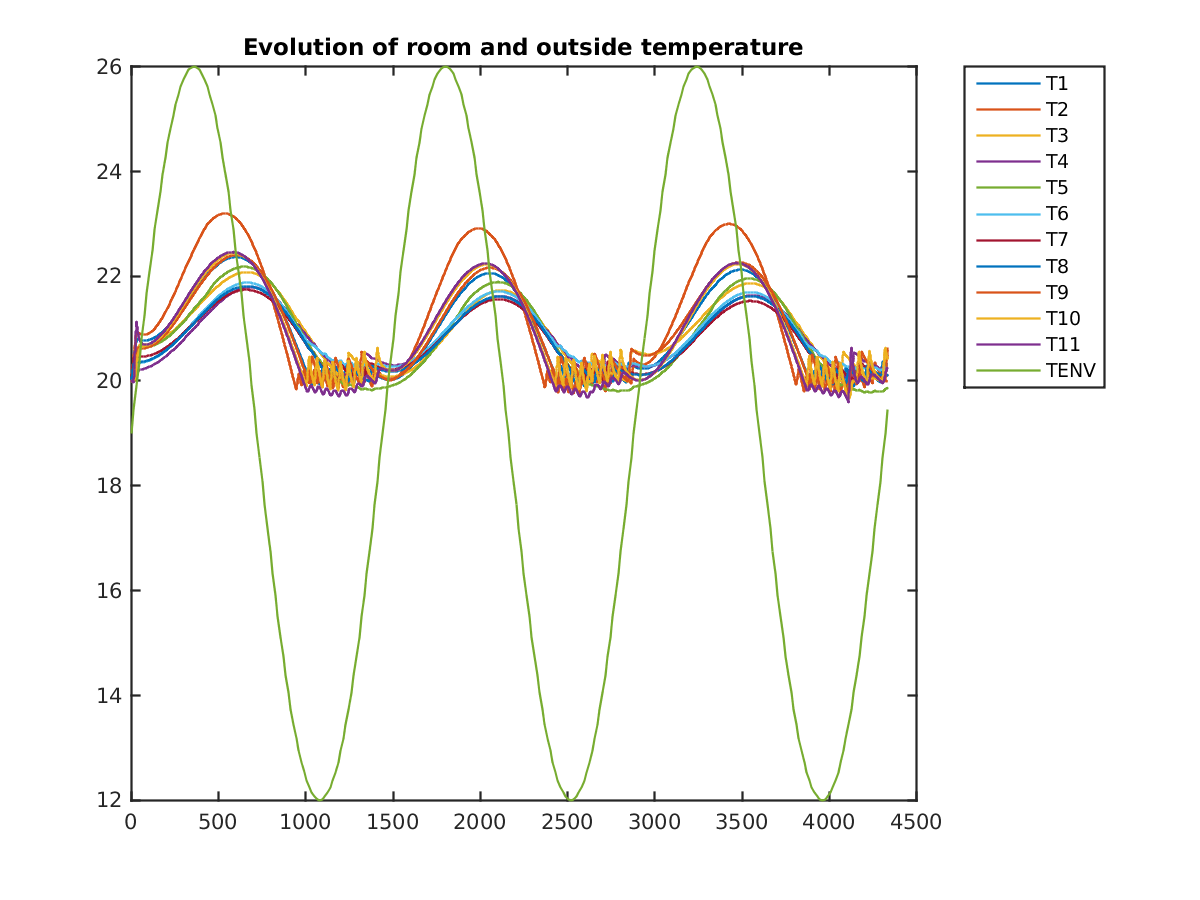}
 \caption{Simulation of the Seluxit case study in the spring scenario.}
 \label{fig:reach_11rooms_stab}
\end{figure}

We now perform the same simulations as in Figure \ref{fig:reach_11rooms_10}, except 
that the environment temperature is not fixed at $10^\circ$C but follows
scenarios of soft winter (Figure~\ref{fig:reach_11rooms}) and spring 
(Figure \ref{fig:reach_11rooms_stab}). The environment temperature is plotted
in green in the figures. The spring scenario is taken from \cite{larsen2015online},
and the soft winter scenario is the winter scenario 
of \cite{larsen2015online} with $5$ additional degrees.
We see that our controller, which is designed for $T_{env} = 10^\circ$C still
satisfies the properties of attainability and stability. These simulations 
are very close those obtained in~\cite{larsen2015online}.

\end{document}